# Robust Ultraviolet to Near-infrared Quantum Emitters in Hexagonal Boron Nitride up to 1100 K

Qing-Hai Tan[1,2,3#], Jia-Min Lai[1,2#], Xue-Lu Liu[1,2], Yong-Zhou Xue[1,2], Xiu-Ming Dou[1,2], Bao-Quan Sun[1,2], Wei-Bo Gao[3], Ping-Heng Tan[1,2,5], Jun Zhang[1,2,4,5]*

[1] *State Key Laboratory of Superlattices and Microstructures, Institute of Semiconductors, Chinese Academy of Sciences, Beijing 100083, China*

[2] *Center of Materials Science and Optoelectronics Engineering, University of Chinese Academy of Sciences, Beijing 100049, China*

[3] *Division of Physics and Applied Physics, School of Physical and Mathematical Sciences, Nanyang Technological University, 637371, Singapore*

[4] *Beijing Academy of Quantum Information Science, Beijing 100193, China*

[5] *CAS Center of Excellence in Topological Quantum Computation, University of Chinese Academy of Sciences, Beijing 101408, China*

[#] These authors contributed equally

**Correspondence and requests for materials should be addressed to J. Z. (Email: zhangjwill@semi.ac.cn)*

## Abstract

A stable single-photon source working at high temperatures with high brightness and covering full band emission from one host material is critically important for quantum technologies. Here, we find that the certain hBN single-photon emissions (SPEs) can be significantly enhanced by lasers with special wavelength, which largely broaden the wavelength range of the hBN emitters, down to ultraviolet (357 nm) and up to near-infrared (912 nm). Importantly, these hBN SPEs are still stable even at the temperature up to 1100 Kelvin. The decoupling between single-photon and acoustic phonon is observed at high temperatures. Our work suggests that hBN can be a good host material for generating single-photon sources with ultrabroad wavelength range.

## Introduction

Solid-state single-photon sources (SPSs) play fundamental roles in quantum technologies including quantum computing, quantum communication, and quantum metrology[1]. Many solid single-photon emitters have been developed such as NV centers in diamond[2], quantum dots[3-6], InGaN[7], as well as emitters in two-dimensional (2D) transition metal dichalcogenide (TMD) materials[8, 9]. SPS at different wavelength ranges has unique applications in quantum technologies. The ultraviolet SPSs can be used for more compact quantum-optical devices[10] while the near-infrared SPSs at telecom wavelength range can be used to realize quantum key distribution and wireless communication[5, 11-13]. Therefore, it is favorable to achieve bright and stable SPEs covering ultraviolet to near-infrared spectrum in a single host material. In 2D hexagonal boron nitride (hBN), bright and stable SPE survives at room temperature and provides the possibility of spin-photon interfaces[14-30]. However, until now, the light-excited SPEs of hBN were found mostly in visible spectral ranges at room temperature[19, 31, 32]. It remains a question whether we can push the wavelength of hBN emitters to ultraviolet and near-infrared.

Here, by using different excitation wavelengths, we observe broadband hBN SPEs covering from ultraviolet (357 nm) to near-infrared (912 nm) at room temperature. Remarkably, these SPEs are stable up to a temperature of 1100 K. The decoupling between photons and acoustic phonons is observed when the temperatures are over 500 K.

## Methods:

***Raman measurement*:** Confocal Raman measurements on hBN samples were undertaken in backscattering geometry with a Jobin-Yvon HR800 system equipped with a liquid-nitrogen-cooled charge-coupled detector. A 100x objective lens (NA=0.9) and 1800 lines $mm^{-1}$ gratings are used for the measurements. The excitation wavelength is 532 nm and the filter is a three-volume Bragg grating filter (532 nm) that can efficiently suppress the Rayleigh signal.

***Photoluminescence (PL) measurement*:** PL spectra measurements were undertaken in backscattering geometry with a Jobin-Yvon HR800 system equipped with a liquid-nitrogen-cooled charge-coupled detector. The samples are cooled by the Montana cryostat system. A 50x long-working-distance objective lens (NA=0.5) and both 600 and 2400 lines $mm^{-1}$ gratings were used for the PL measurements at low temperatures. The highest resolution of our system with 2400 lines $mm^{-1}$ grating is around 40 μeV.

***The second-order correlation function measurement*:** The second-order correlation function measurement is carried out by using a home-built Hanbury-Brown-Twiss (HBT) setup. Two silicon Avalanche Photo Diodes (APD) are used to count photons.

## Results and Discussion

### Observation of single-photon emission in hBN

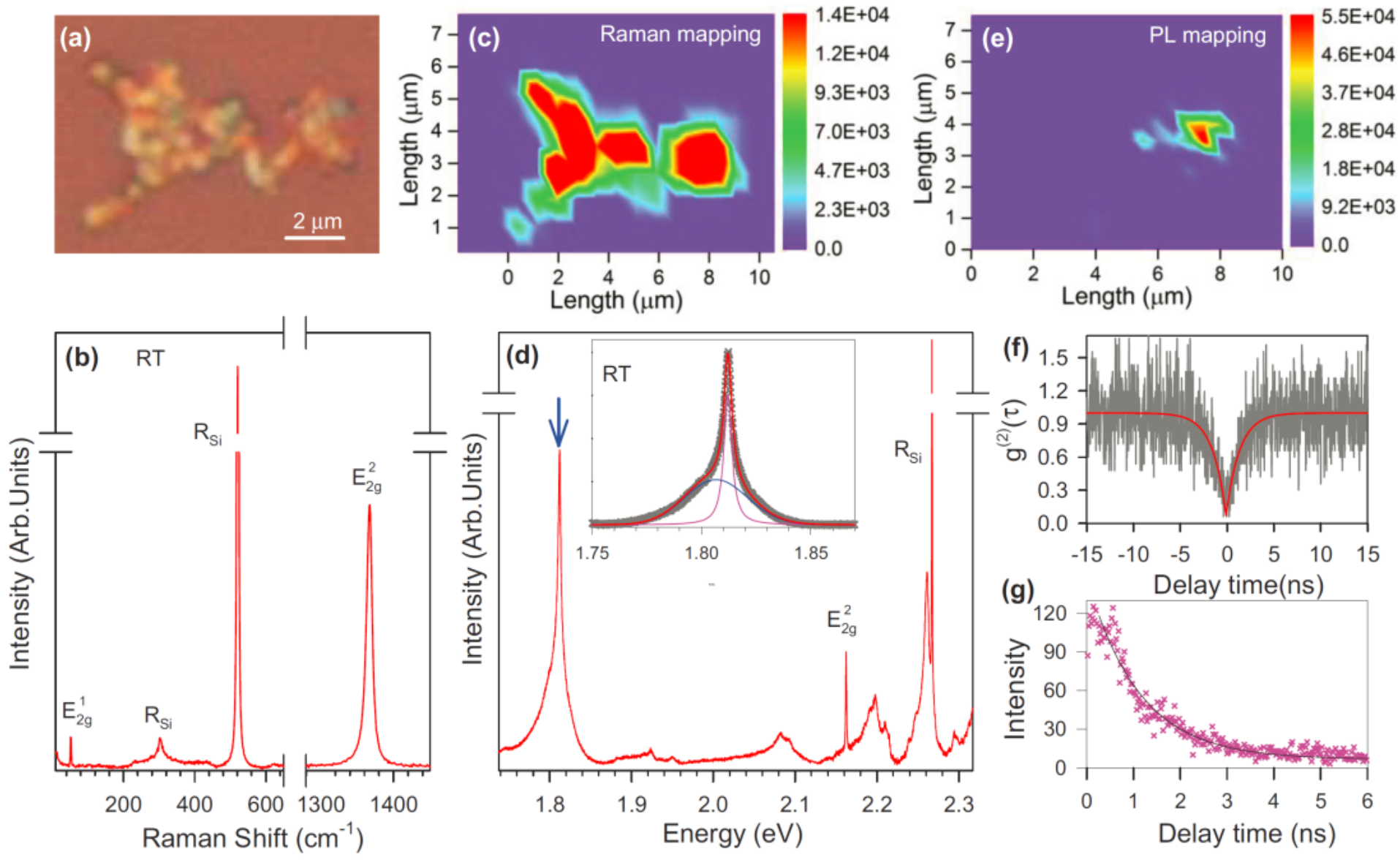

**Fig. 1** | (a) The optical microscopy image of hBN sample. (b) Raman spectrum of hBN sample in (a). (c) Confocal Raman mapping of $E^2_{2g}$ mode over the same area in (a). (d) PL spectrum of hBN SPE in (a) at room temperature, the inset shows the fitting results of the PL peak with two components, i.e., zero phonon line and phonon sideband. (e) Confocal PL mapping with a bandpass at 1.81 eV. (f) The second-order correlation function measurement of PL peak in (d). (g) Time-resolved intensity of PL peak in (d).

The hBN samples are prepared on $SiO_2$/Si substrate from commercial hBN flakes suspended in 50/50 ethanol/water solution (Graphene Supermarket). After dispersing 20 μL solution on $SiO_2$/Si substrate, we dried up the samples by using a warm white light. Figure 1(a) is a typical optical microscopic image of the prepared sample. Fig. 1(b) shows the Raman spectrum of the hBN sample. The detailed methods for optical measurements are present in the Experimental Methods. Two Raman modes of $E^1_{2g}$ at 52 cm$^{-1}$ (FWHM= 2.93 cm$^{-1}$) and $E^2_{2g}$ at 1366.6 cm$^{-1}$ (FWHM=10.4 cm$^{-1}$) are observed, consistent with the Raman spectra of exfoliated multilayer hBN in previous reports[33, 34]. Figure 1(c) shows the corresponding confocal Raman mapping ($E^2_{2g}$ mode) of hBN at room temperature, which matches the microscopic image of the hBN sample as shown in Fig. 1(a), confirming the crystal uniformity of the sample.

Figure 1(d) shows the room-temperature photoluminescence (PL) spectra taken at a certain spot in Fig. 1(a). Besides the Raman modes of the silicon substrate and hBN, we also observed a bright PL peak around 1.81 eV (~684 nm), similar to the previously reported SPEs in hBN[17, 19, 35]. The inset in Fig. 1(d) shows the fitting results of this PL peak. The narrow one with a Lorentzian shape (~4.4 meV linewidth) is the ZPL and the other broader feature (~39 meV linewidth) with a Gaussian shape is the phonon sideband (PSB)[23]. The fitting results suggest the ZPL here is dominated by the homogeneous broadening mechanism induced by the thermal phonon interaction[24, 36, 37]. Because the PSB represents a Poisson distribution composed of electronic transitions with a discrete number of phonons, it shows as a Gaussian line shape[38]. We estimated the Debye-Waller factor (defined as the integrated intensity of ZPL divided by the integrated PL) to be around 0.65, suggesting the ZPL dominates the spectrum. In contrast to the Raman mapping of the $E_{2g}^{2}$ mode, the PL mapping of the emission peak at 1.81 eV only exists in certain location of the sample (see Fig. 1(e)), implying localized feature of the corresponding exciton. Fig. 1(f) is the measured second-order correlation function ($g^{(2)}(\tau) = 1 - ae^{(-|\tau|/\tau_0)}$, where $a$ is the background of uncorrelated photons and $\tau_0$ is the antibunching recovery time) of the peak at 1.81 eV. The $g^{(2)}(0)$ (~0.09) is below 0.5, confirming the single-photon of the emitter. The extracted lifetime ($\tau_0$) from $g^{(2)}(\tau)$ is around 1.4 ns, consistent with the time-resolved PL result in Fig.1 (g).

**Ultraviolet to near-infrared hBN quantum emitters at room temperature**

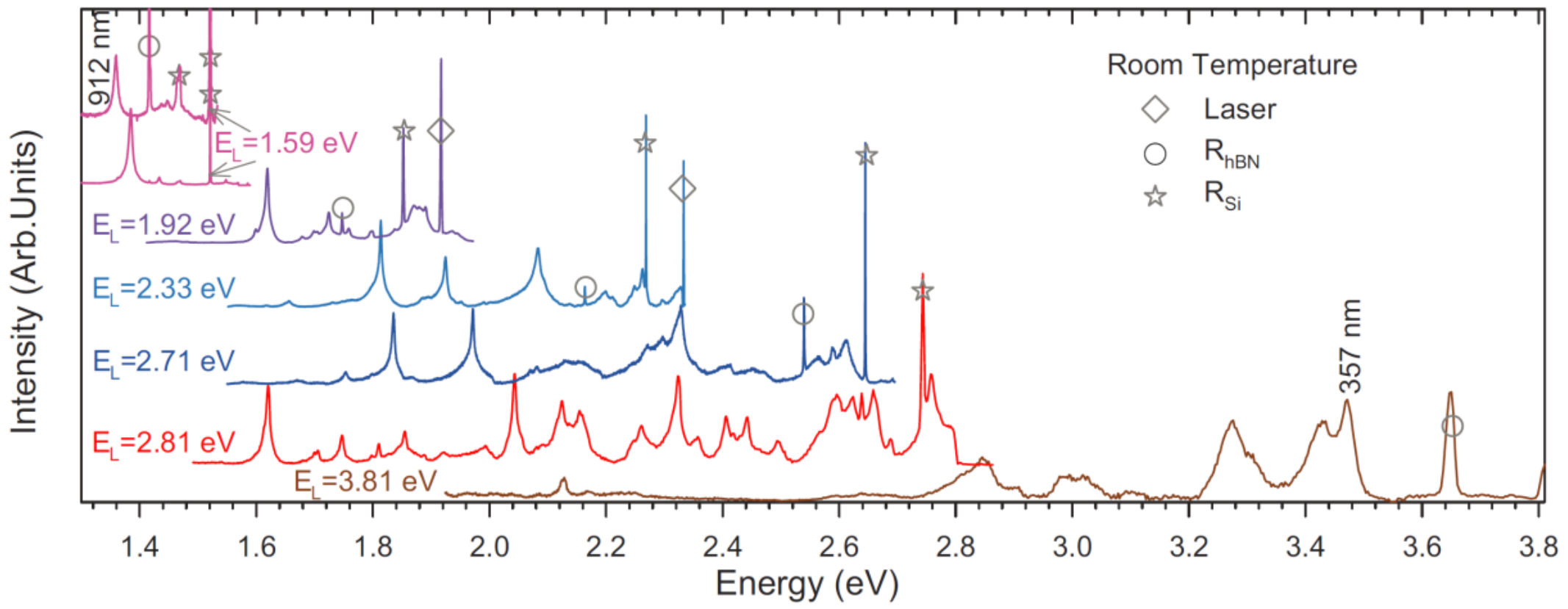

**Fig. 2** | PL spectra of different hBN samples with six different excitation wavelengths at room temperature. The star, circle, and diamond symbols mark the Raman mode of silicon, $E_{2g}^{2}$ mode of hBN and the excitation laser lines, respectively.

We measured many samples with different excitation wavelengths ranging from ultraviolet to near-infrared, as shown in Fig. 2. We observed a series of narrow PL peaks from 357 nm to 912 nm at room temperature, much broader than previously reported results that mainly span from 550 nm to 800 nm[19, 36, 39]. These results suggest ultraviolet to near-infrared single-photon emitters can be achieved by selecting special excitation wavelengths, which implies that there might be multiple defect levels or types in our hBN sample. These defects can be selectively excited, emitting SPE at different wavelengths. Compared with the predicted results by theory calculations[33, 34, 40, 41], these SPE peaks may originate from defect levels related to vacancy, carbon, and oxygen, such as $V_NC_B$ (~2.5 eV), $V_NN_B$ (~2 eV), $O_B$ (~1.5 eV).

To study the stability of these SPEs, we detected the temporal evolution and power dependency of these SPEs, as shown in Fig. 3. Most of these SPEs are stable, bright with the temporal evolution from 0-80 s (Fig. 3(a)), and saturated at higher power, as shown in Fig 3(b). Our results indicate that if the excitation conditions match well with the infrared defect levels in hBN, the communication band SPEs based on hBN is expected to be observed at room temperature.

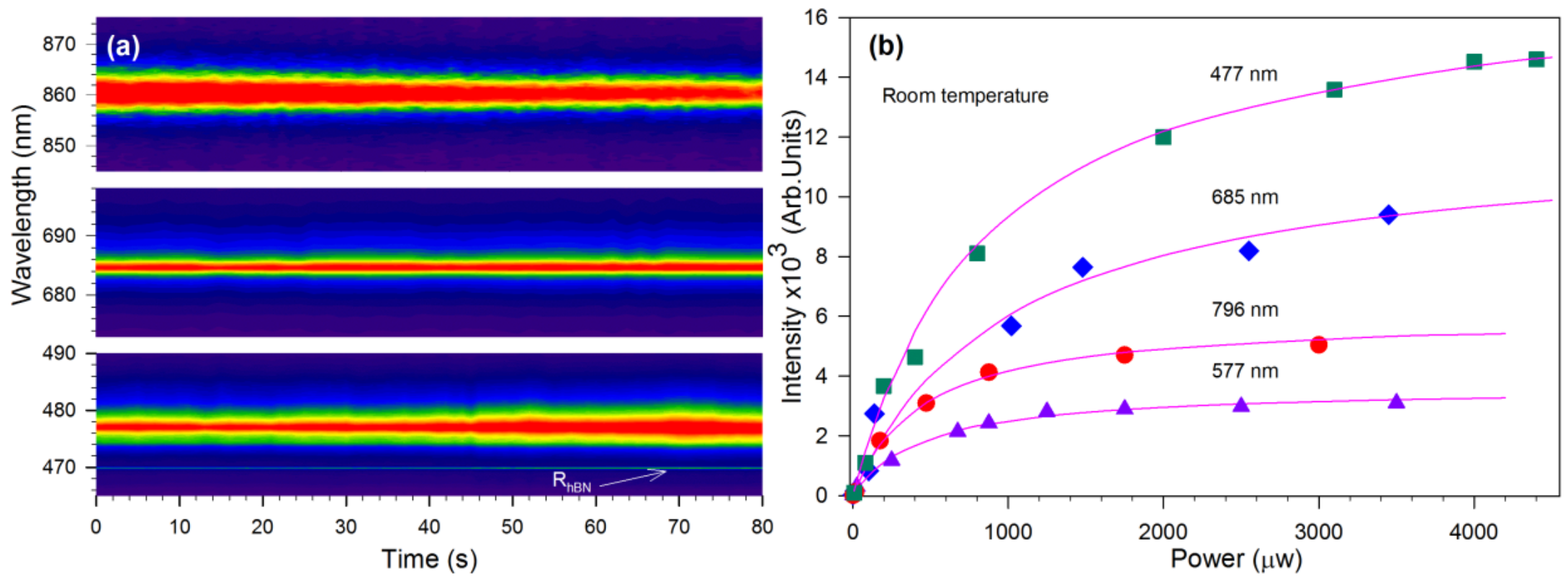

**Fig. 3** | (a) The temporal evolution of three different defect emitters under continuous illumination at room temperature. (b) The PL intensity of emitters centered at 477, 577, 685, and 796 nm from the different positions as a function of laser power at room temperature, respectively.

**High temperatures hBN SPE up to 1100 K**

To study the temperature stability of hBN SPEs, we measured the temperature dependence of hBN SPEs from 4K to 1100 K, as shown in Fig. 4. Surprisingly, these SPE peaks can survive even at the temperature increases to 1100 K. We should note that due to the limits of our equipment, the actual limit temperature of stable hBN SPEs may be higher than 1100 K. The stable hBN SPEs at high temperatures can broaden their application under extremes conditions with high temperature. To further reveal the temperature dependence of the ZPL position and linewidth, we fitted one representative emission peak in Fig. 4(a) illustrated by an arrow, and plotted the extracted parameters in Fig. 4(b). We also show the temperature dependence of the ZPL position and linewidth from 4 K to 300 K as a contrast (Fig. 4(c)). From 4 K to 300 K, the linewidths show a $T^3$ temperature dependence, consistent with the previous results. The ZPL positions can be well described by using O'Donnell equation[42]. We found the energy of PL ZPL exhibits a linear temperature dependence, which can be understood by considering the electron-phonon interactions and the lattice expansions over temperatures[18, 43-45].

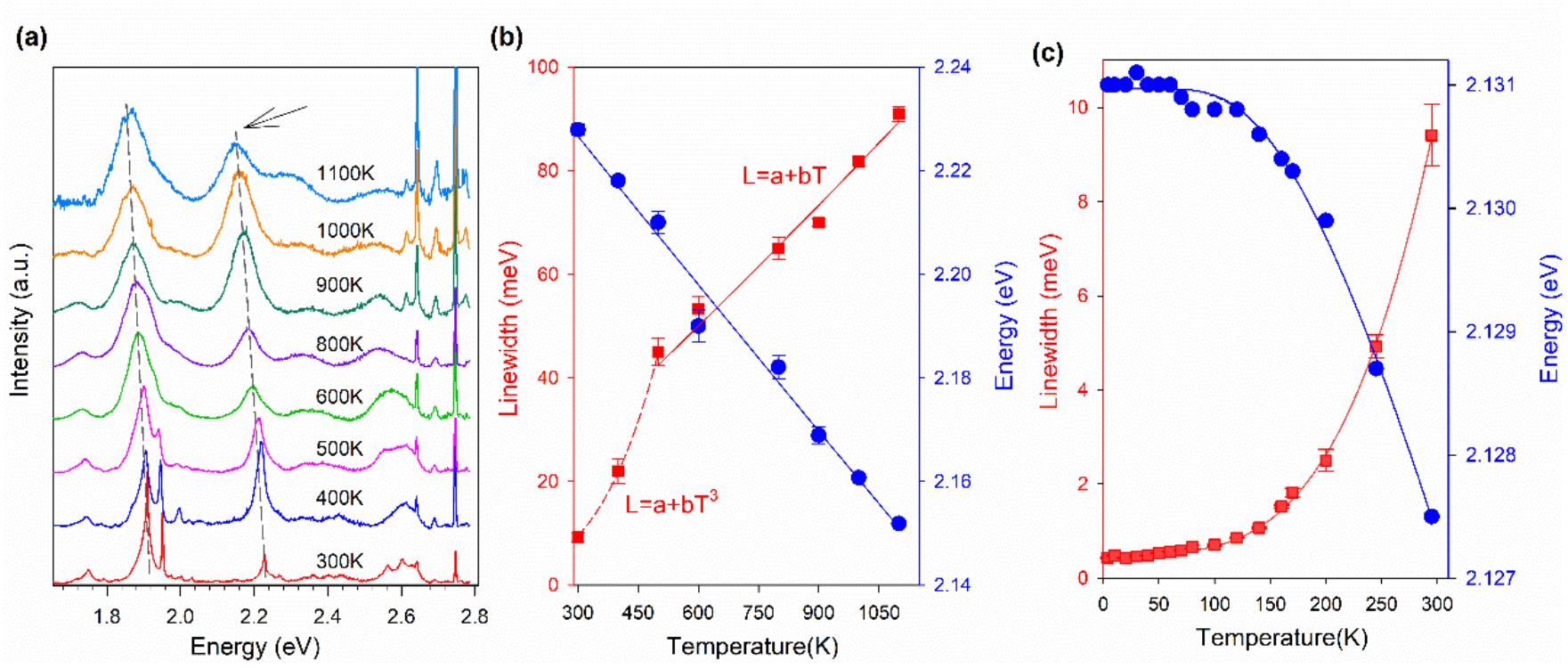

**Fig. 4** | (a) Temperature dependence of hBN SPEs from 300 K to 1100 K. (b) Temperature dependence of linewidth and energy of the peak illustrated by arrows in (a) from 300 K to 1100 K. The red line in (b) is fitted by the function $at^{\alpha} + c$, where $\alpha$=3.54 (±0.31) for dash line in the temperatures range from 300 K to 500 K, where $\alpha$=1.11 (±0.06) for solid line in the temperatures range from 500 K to 1100 K. The solid blue line in (b) is linear fits of peak energy. (c) Temperature dependence of linewidth and energy of a peak at 2.131 eV (4 K) from 4 K to 300 K. The red solid line in (c) is fitted by the function $aT^{\alpha} + b$, where $a = 3.75\ (0.10) \times 10^{-10}$, $\alpha$=3.00 (0.20) and $b = 3.45\ (0.70) \times 10^{-4}$ eV. The energy of ZPL is fitted by the O'Donnel equation (blue line) $E_g(T)= E_g(0) - S<\hbar\omega>$, where $S$ is an exciton-phonon coupling constant and $<\hbar\omega>$ is the average phonon energy. Here $E_g(0)$, $S$ and $<\hbar\omega>$ is around 2.14 eV, 0.23 and 53.25 meV, respectively.

Generally, at low temperatures, the ZPL is dominated by the inhomogeneous broadening induced by spectral diffusion, exhibiting as a Gaussian profile; as temperature increases, homogeneous broadening induced by phonons starts to take a dominant role, leading to a Lorentzian profile at higher temperatures[37, 46, 47]. Interestingly, we found the temperature dependence of hBN SPEs linewidth shows an unusual behavior. Specifically, for temperatures ranging from 300 K to 500 K, the linewidths show a $T^3$ temperature dependence, consistent with the results from 4 K to room temperature (see Fig. 4(c)) and the previous results[18, 38, 39, 45, 46]; while for temperatures above 500 K, the linewidths show a linear response. This temperature dependence behavior may be understood by considering the Debye temperature of hBN ($T_D$~410 K for bulk hBN[47]). For temperatures below $T_D$, phonon-induced homogeneous broadening dominates the linewidth and leads to $T^3$ dependence[18, 48, 49]. Above $T_D$, the linear temperature response can be explained by considering the decoupling between photon and acoustic phonon[45], consistent with the Gaussian profile of ZPL at high temperature (T>500 K). To thoroughly figure out the linewidth broadening mechanism and the phonons' role for hBN SPEs, more investigations are required.

## Conclusions

In conclusion, we observed ultraviolet (357 nm) to near-infrared (912 nm) ultrabroad-band single-photon emitters (SPEs) in hexagonal boron nitride (hBN) by varying the excitation wavelength. These single-photon sources can be operated from room temperature to 1100 K. Our studies suggest that hBN could be an ideal host material for generating broadband SPEs at higher temperatures.

**Conflict of interest** The authors declare no conflict of interest.

**Acknowledgments** Z.J. acknowledges support from National Basic Research Program of China (grant no. 2017YFA0303401), CAS Interdisciplinary Innovation Team, Strategic Priority Research Program of Chinese Academy of Sciences (grant no. XDB28000000). W.G. acknowledges Singapore National Research Foundation and DSO National Laboratories under the QEP grant NRF2021-QEP2-03-P01, 2019-0643 (QEP-P2), 2019-1321 (QEP-P3), CRP Award No. NRF-CRP21-2018-0007, NRF-CRP22-2019-0004, and NRF-CRP23-2019-0002); Singapore Ministry of Education (MOE2016-T3-1-006 (S)).

**Author contributions** J.Z. and Q.T. conceived the ideas; Q.T., J.Z., P.T., and S.B designed the experiments. Q.T., J.L. and X.L. prepared the samples. Q.T., J.L., X.L., Y.X., and X.D. performed experiments. Q.T., J.L. W.G., and Z.J. analyzed the data and wrote the manuscript with inputs from all authors.

**Additional information**

Reprints and permissions information is available online.

Correspondence and requests for materials should be addressed to J.Z. (Email: zhangjwill@semi.ac.cn).